\documentclass[12pt]{article}
\usepackage{epsfig}
\begin{document}
\title{Near-threshold $J/\Psi$ production in proton--nucleus collisions}
\author{Yu. T. Kiselev$^1$, E. Ya. Paryev$^{1,2}$, and Yu. M. Zaitsev$^1$\\
{\it $^1$Institute for Theoretical and Experimental Physics,}\\
{\it Moscow 117218, Russia}\\
{\it $^2$Institute for Nuclear Research, Russian Academy of Sciences,}\\
{\it Moscow 117312, Russia}}

\renewcommand{\today}{}
\maketitle

\begin{abstract}
   We study the $J/\Psi$ production from nuclei near the kinematic threshold within
   the collision model, based on the nuclear spectral function,
   for incoherent primary proton--nucleon charmonium creation processes.
   The model takes into account the initial proton and final $J/\Psi$ absorption, target nucleon
   binding and Fermi motion, the formation length of $J/\Psi$ mesons as well as the effect of their
   nuclear mean-field potential on these processes. We calculate the A dependences of the absolute and
   relative (transparency ratio) charmonium yields as well as its excitation function within the different
   scenarios for the $J/\Psi N$ absorption cross section, for the $J/\Psi$ formation length and for $J/\Psi$ in-medium
   modification. We demonstrate that the former observables, on the one hand, are not practically affected
   by the charmonium formation length effects and, on the other hand, they are appreciably sensitive to the genuine
   $J/\Psi N$ absorption cross section at beam energies of interest, which means that these observables can
   be useful to help determine the above cross section from the comparison of the results of our calculations
   with the future data from the CBM experiment at FAIR--the upcoming accelerator facility at GSI-Darmstadt,
   Germany. We also show that the excitation function for $J/\Psi$ subthreshold production in $pA$ reactions
   reveals some sensitivity to adopted its in-medium modicication scenarios. Therefore, such
   observable may be an important tool to get valuable information on the charmonium in-medium
   properties.
\end{abstract}

\newpage

\section*{1. Introduction}

\hspace{1.5cm} An extensive investigations of the charmonium production and suppression in relativistic
proton--nucleus and nucleus--nucleus collisions have been carried out over the last years
(see, for example, [1--11]). From these studies one hopes to extract valuable information about
both the strength of the inelastic $J/\Psi$--nucleon interaction and a possible formation of a
quark-gluon plasma in these collisions, in-medium properties of charmonium states. However, in
high-energy collisions the $J/\Psi$ mesons are produced with high momenta relative to the nuclear medium,
which does not allow one to put strong constraints on the genuine charmonium--nucleon absorption cross
section from the existing experimental data due to the substantial $J/\Psi$ formation time effects.
There are essentially less extensive investigations [12--17] of the production and absorption of the
charmonium states in antiproton--nucleus reactions at low energies, where they are created with the moderate
momenta relative to the target nucleus and where, therefore, the $J/\Psi$ formation length effects play a
minor role. Finally, the production of $J/\Psi$ mesons on nuclei in proton-induced reactions
at low energies [18--20] and, especially, at energies
close to their kinematic threshold (the threshold kinetic energy for $J/\Psi$ production in $pp$ collisions
being 11.3 GeV) [11, 21] has up to now received very little consideration, probably, because of the lack of
suitable facilities and associated detectors. Since the cross sections of ($p, J/\Psi$) reaction at
near-threshold energies are expected to be extremely low, high intensity proton beams are needed to perform
accurate measurements of $J/\Psi$ creation on nuclear targets in the threshold region. Such measurements
are planned to be conducted in the near future in the upcoming compressed baryonic matter (CBM) experiment
at the Facility for Anti-proton and Ion Research (FAIR).

The main aim of the present work is to get the estimates of the absolute and relative yields of $J/\Psi$
mesons and their excitation functions from $pA$ collisions in the near-threshold energy region in the framework
of the first collision model based on the nuclear spectral function. In view of the future data from FAIR
experiments, these estimates can be used as an important tool for understanding the role of the conventional
cold nuclear matter effects in $J/\Psi$ production in nuclear collisions at various beam energies.

\section*{2. First collision model}

\hspace{1.5cm} Since we are interested in the near-threshold bombarding energy region up to
14 GeV, we have taken into consideration the following elementary processes,
which have the lowest free production threshold ($\approx$ 11.3 GeV)
\footnote{$^)$We can neglect in the energy domain of our interest the following two-step $J/\Psi$ production
processes with $\chi_{c1}$, $\chi_{c2}$ and $\Psi^\prime$ mesons in an intermediate states:
 $pN \to pN\chi_{c1}$, $pN \to pN\chi_{c2}$, $pN \to pN\Psi^\prime$; $\chi_{c1} \to J/\Psi\gamma$
 (${\rm BR}=36$\%), $\chi_{c2} \to J/\Psi\gamma$ (${\rm BR}=20$\%),
 $\Psi^\prime \to J/\Psi{\pi}{\pi}$ (${\rm BR}=49$\%) and $\Psi^\prime N \to J/\Psi N$ due to larger
$\chi_{c1}$, $\chi_{c2}$ and $\Psi^\prime$ production thresholds in $pN$ collisions--13.6, 13.9 and
14.6 GeV, respectively.}$^)$
:
\begin{equation}
p+p \to p+p+J/\Psi,
\end{equation}
\begin{equation}
p+n \to p+n+J/\Psi.
\end{equation}
It is expected [22--26] that the $J/\Psi$ mass shift at saturation density $\rho_0$ is only
of the order of a few MeV due to a small coupling of the $c$, $\bar c$ quarks to the nuclear medium.
However, in Ref. [27] Sibirtsev and Voloshin have shown within the multipole expansion and low-energy
theorems in QCD that this mass shift is significantly larger ($-{\Delta}m_{J/\Psi} \ge 21$ MeV) than
previously estimated in the literature and might be observed experimentally. So, it is unclear currently
what $J/\Psi$ in-medium mass shift is the correct one. The knowledge of this shift is important for
distinguishing between different models [22--27] and, hence, for gaining more insight into the dynamics
of low-energy QCD.
Accounting for mentioned above, in the following calculations we will include
the medium modification of the $J/\Psi$ mesons, participating in the production processes (1),
(2), by using, for reasons of simplicity, their average in-medium mass $<m^*_{{J/\Psi}}>$ defined as:
\begin{equation}
<m^*_{{J/\Psi}}>=\int d^3r{\rho_N({\bf r})}m^*_{{J/\Psi}}({\bf r})/A,
\end{equation}
where ${\rho_N({\bf r})}$ and $m^*_{{J/\Psi}}({\bf r})$ are the local nucleon density and
${J/\Psi}$ effective mass inside the nucleus, respectively. Assuming that
\begin{equation}
m^*_{{J/\Psi}}({\bf r})=m_{{J/\Psi}}+V_0\frac{{\rho_N({\bf r})}}{{\rho_0}},
\end{equation}
we can readily rewrite equation (3) in the form
\begin{equation}
<m^*_{{J/\Psi}}>=m_{{J/\Psi}}+V_0\frac{<{\rho_N}>}{{\rho_0}}.
\end{equation}
Here, $m_{{J/\Psi}}$ is the ${J/\Psi}$ free space mass and $<{\rho_N}>$ is the
average nucleon density. Our calculations show that, for example, for target nuclei $^{12}$C
and $^{93}$Nb the ratio $<{\rho_N}>/{\rho_0}$ is approximately equal to 0.5 and 0.7, respectively.
We will use these values throughout the following study. In it for the $J/\Psi$ mass shift at
saturation density $V_0$ we will employ the four following options: i) $V_0=0$, ii) $V_0=-50$ MeV,
iii) $V_0=-100$ MeV, and iv) $V_0=-150$ MeV. For the reason of reducing the possible uncertainty of our
calculations due to the use in them of the model nucleon [28, 29] self-energies at high momenta, we will
ignore the medium modification of the outgoing nucleon mass in the present work.

  Then, taking into account the distortion of the incident proton and the absorption of the final
${J/\Psi}$ meson as well as the fact that in the near-threshold energy region of our interest
the produced $J/\Psi$ meson moves in the nucleus substantially in the forward direction and using the
results given in [16, 30, 31], we can represent the total
\footnote{$^)$In the full phase space without any cuts on angle and momentum of
the observed ${J/\Psi}$ meson.}$^)$ cross section for the production of ${J/\Psi}$ mesons
off nuclei in the primary proton--induced reaction channels (1), (2) as follows:
\begin{equation}
\sigma_{pA\to {J/\Psi}X}^{({\rm prim})}(T_0)=I_{V}[A]
\left<\sigma_{pN \to pN{J/\Psi}}(T_0)\right>_A,
\end{equation}
where
\begin{equation}
I_{V}[A]=2{\pi}A\int\limits_{0}^{R}r_{\bot}dr_{\bot}
\int\limits_{-\sqrt{R^2-r_{\bot}^2}}^{\sqrt{R^2-r_{\bot}^2}}dz
\rho(\sqrt{r_{\bot}^2+z^2})
\end{equation}
$$
\times
\exp{\left[-\sigma_{pN}^{\rm in}A\int\limits_{-\sqrt{R^2-r_{\bot}^2}}^{z}\rho(\sqrt{r_{\bot}^2+x^2})dx
-A\int\limits_{z}^{\sqrt{R^2-r_{\bot}^2}}\sigma_{{J/\Psi}N}^{\rm eff}(x-z)
\rho(\sqrt{r_{\bot}^2+x^2})dx\right]},
$$
\begin{equation}
\left<\sigma_{pN \to pN{J/\Psi}}(T_0)\right>_A=
\int\int
P_A({\bf p}_t,E)d{\bf p}_tdE
\sigma_{pN \to pN{J/\Psi}}(\sqrt{s},<m^*_{{J/\Psi}}>)
\end{equation}
and
\begin{equation}
  s=(E_{0}+E_t)^2-({\bf p}_{0}+{\bf p}_t)^2,
\end{equation}
\begin{equation}
   E_t=M_A-\sqrt{(-{\bf p}_t)^2+(M_{A}-m_{N}+E)^{2}}.
\end{equation}
Here,
$\sigma_{pN\to pN{J/\Psi}}(\sqrt{s},<m^*_{{J/\Psi}}>)$ is the "in-medium"
total cross section for the production of ${J/\Psi}$
with reduced mass $<m^*_{{J/\Psi}}>$ in reactions (1) and (2)
\footnote{$^)$In equation (6) it is assumed that the $J/\Psi$ meson production cross sections
in $pp$ and $pn$ interactions are the same.}$^)$ at the $pN$ center-of-mass energy $\sqrt{s}$;
$\rho({\bf r})$ and $P_A({\bf p}_t,E)$ are the local nucleon density and the
spectral function of target nucleus $A$ normalized to unity
\footnote{$^)$The specific information about used in our subsequent calculations these quantities
is given in [30, 31].}$^)$;
${\bf p}_{t}$  and $E$ are the internal momentum and binding energy of the struck target nucleon
just before the collision; $\sigma_{pN}^{\rm in}$ is the inelastic cross section
\footnote{$^)$We use $\sigma_{pN}^{\rm in}=30$ mb in our present calculations.}$^)$
of the free pN interaction; $A$ is the number of nucleons in
the target nucleus, $M_{A}$  and $R$ are its mass and radius; $m_N$ is the bare nucleon mass;
${\bf p}_{0}$, $E_0$ and $T_0$ are the momentum, total and kinetic energies of the initial proton;
$\sigma_{{J/\Psi}N}^{\rm eff}(z)$ is the $J/\Psi$--nucleon effective absorption cross section, which
will be defined below. The quantity $I_{V}[A]$ in equation (6) represents the effective number of target
nucleons participating in the primary $pN\to pN{J/\Psi}$ reactions.

  Following [31, 32], we assume that the "in-medium" cross section
$\sigma_{pN \to pN{J/\Psi}}({\sqrt{s}},<m^*_{J/\Psi}>)$ for $J/\Psi$ production in reactions (1) and (2)
is equivalent to the vacuum cross section $\sigma_{pN \to pN{J/\Psi}}({\sqrt{s}},m_{J/\Psi})$ in which
the free mass $m_{J/\Psi}$ is replaced by the average in-medium mass $<m^*_{{J/\Psi}}>$ as given by
equation (5). For the free total cross section
$\sigma_{pN \to pN {J/\Psi}}({\sqrt{s}},m_{J/\Psi})$ we have used the parametrization
\footnote{$^)$It should be pointed out that this parametrization describes the inclusive
$pp \to J/\Psi X$ cross section which is assumed to be the same as that for $pN \to pN J/\Psi$ at beam
energies of interest.}$^)$
from [3, 5] that has been corrected for the proper threshold behavior with accounting for
the lowest data point from [19] taken at 24 GeV/c incident proton momentum, viz.:
\begin{equation}
\sigma_{pN \to pN{J/\Psi}}({\sqrt{s}},m_{J/\Psi})=\left\{
\begin{array}{ll}
	\frac{2{\cdot}37}{0.0597}\left(1-\frac{m_{J/\Psi}}{\sqrt{s}}\right)^{12}~[{\rm nb}]
	&\mbox{for $\sqrt{s} > 10~{\rm GeV}$}, \\
	&\\
                   528\left(1-\frac{m_{J/\Psi}+2m_N}{\sqrt{s}}\right)^{5.225}~[{\rm nb}]
	&\mbox{for $m_{J/\Psi}+2m_N \le \sqrt{s} \le 10~{\rm GeV}$}.
\end{array}
\right.	
\end{equation}

 Let us focus now on the charmonium--nucleon effective cross section
$\sigma_{{J/\Psi}N}^{\rm eff}(z)$, entering into the equation (7), which takes into account the
time dependence of the $J/\Psi$ formation. Following [13, 16, 33], we express this cross section in
terms of a $J/\Psi$ meson formation length $l_{J/\Psi}$:
\begin{equation}
\sigma_{{J/\Psi}N}^{\rm eff}(z)=\sigma_{{J/\Psi}N}\left\{\theta(l_{J/\Psi}-z)\left[\frac{z}{l_{J/\Psi}}+
\frac{n^2<k^2_t>}{m^2_{J/\Psi}}\left(1-\frac{z}{l_{J/\Psi}}\right)\right]+\theta(z-l_{J/\Psi})\right\}.
\end{equation}
Here, $\sigma_{{J/\Psi}N}$ is the genuine free-space $J/\Psi$--nucleon absorption cross section;
n is the number of valence quarks of the hadron ($n=2$ in our case), while $<k^2_t>^{1/2}$ is the
average transverse momentum of the quark in the hadron (taken to be $<k^2_t>^{1/2}=0.35$ GeV/c);
$z$ is the distance from the $c\bar{c}$-pair production point and $\theta(x)$ is the standard step function.
For the $J/\Psi$ meson formation length $l_{J/\Psi}$ we adopt the conventional formula with an energy
denominator [13, 16, 33]:
\begin{equation}
l_{J/\Psi}\simeq\frac{2p^{\rm lab}_{J/\Psi}}{m^2_{\Psi^\prime}-m^2_{J/\Psi}},
\end{equation}
where $p^{\rm lab}_{J/\Psi}$ is the charmonium momentum in the target nucleus rest frame.
Taking into consideration that the momentum $p^{\rm lab}_{J/\Psi}\simeq{7.6}$ GeV/c in the process
$pN \to pN{J/\Psi}$ occuring on a free target nucleon being at rest at threshold energy
of 11.3 GeV, one can get that $l_{J/\Psi}\simeq{0.8}$ fm for this momentum. We will use this value for the
quantity $l_{J/\Psi}$ throughout our calculations for a collection of $A$ target nucleons subject to
Fermi motion in the near-threshold energy domain. As is easy to see
from formula (12), when $J/\Psi$ formation length $l_{J/\Psi} \to 0$ then the effective cross section
$\sigma_{{J/\Psi}N}^{\rm eff}(z)\to \sigma_{{J/\Psi}N}$, which means that in this case the "normal"
$J/\Psi$ absorption in nuclear matter is recovered. It should be noticed that the second exponent in
equation (7) can be put in view of equation (12) in the following in an easy-to-use in numerical integration
form:
\begin{equation}
A\int\limits_{z}^{\sqrt{R^2-r_{\bot}^2}}\sigma_{{J/\Psi}N}^{\rm eff}(x-z)
\rho(\sqrt{r_{\bot}^2+x^2})dx=\sigma_{{J/\Psi}N}A\theta(\sqrt{R^2-r_{\bot}^2}-z-l_{J/\Psi})
\end{equation}
$$
\times
\left\{\int\limits_{z}^{l_{J/\Psi}+z}\left[\frac{x-z}{l_{J/\Psi}}+\alpha(n)\left(1-\frac{x-z}{l_{J/\Psi}}
\right)\right]\rho(\sqrt{r_{\bot}^2+x^2})dx+\int\limits_{l_{J/\Psi}+z}^{\sqrt{R^2-r_{\bot}^2}}
\rho(\sqrt{r_{\bot}^2+x^2})dx\right\}
$$
$$
+
\sigma_{{J/\Psi}N}A\theta(l_{J/\Psi}+z-\sqrt{R^2-r_{\bot}^2})
\int\limits_{z}^{\sqrt{R^2-r_{\bot}^2}}\left[\frac{x-z}{l_{J/\Psi}}+\alpha(n)\left(1-\frac{x-z}{l_{J/\Psi}}
\right)\right]\rho(\sqrt{r_{\bot}^2+x^2})dx,
$$
where $\alpha(n)=n^2<k^2_t>/m^2_{J/\Psi}$ ($\alpha(2)=0.0511$).

     The absorption cross section $\sigma_{{J/\Psi}N}$ can be extracted, in particular, from
a comparison of the calculations with the measured transparency ratio of the $J/\Psi$ meson,
normalized, for example, to carbon:
\begin{equation}
T_A=\frac{12~\sigma_{pA \to {J/\Psi}X}}{A~\sigma_{pC \to {J/\Psi}X}}.
\end{equation}
Here, $\sigma_{pA \to {J/\Psi}X}$ and
$\sigma_{pC \to {J/\Psi}X}$ are inclusive total cross sections for $J/\Psi$ production in
$pA$ and $p{\rm C}$ collisions, respectively. If the primary proton--induced reaction channels
(1), (2) dominate in the $J/\Psi$ production in $pA$ reactions close to threshold
\footnote{$^)$One may expect that this is so due to the following. The main inelastic channel in $pN$
collisions at beam energies of interest is the multiplicity production of pions with comparatively
low energies at which the secondary ${\pi}N \to {J/\Psi}X$ processes are energetically suppressed.}
$^)$
, then, according to (6) and (7), we have:
\begin{equation}
T_A=\frac{12~I_V[A]}{A~I_V[C]}
\frac{\left<\sigma_{pN \to pN{J/\Psi}}(T_0)\right>_A}
{\left<\sigma_{pN \to pN{J/\Psi}}(T_0)\right>_C}.
\end{equation}
Ignoring the medium effects
\footnote{$^)$These effects lead to only small corrections to the ratio $T_A$ at above threshold
incident energies.
They are within several percent here, as our calculations by (16) and (17) for the nucleus with a
diffuse boundary showed. However, the medium effects become substantial at subthreshold beam energies,
as our calculations also demonstrated.}$^)$,
from (16) we approximately obtain:
\begin{equation}
T_A \approx \frac{12~I_V[A]}{A~I_V[C]}.
\end{equation}
The integral (7) for $I_V[A]$ in the case of a nucleus of a radius $R=r_0A^{1/3}$ with a sharp
boundary and in the limit $l_{J/\Psi} \to 0$ has the following simple form [34]:
\begin{equation}
I_V[A]=\frac{3A}{(a_1-a_2)a_2^2}\left\{1-(1+a_2)e^{-a_2}-{\left(\frac{a_2}{a_1}\right)}^2
[1-(1+a_1)e^{-a_1}]\right\},
\end{equation}
where $a_1=3A\sigma_{J/\Psi N}/2{\pi}R^2$ and $a_2=3A\sigma_{pN}^{\rm in}/2{\pi}R^2$.
The simple formulas (17), (18) allow one to easily estimate the transparency ratio $T_A$
at well above threshold energies.

  Let us discuss now the results of our calculations in the framework of the approach outlined above.

\section*{3. Results}

\hspace{1.5cm} Figure 1 shows the A--dependence of the total $J/\Psi$ production
cross section from the primary $pN \to pN{J/\Psi}$ reaction channels
in $pA$ ($A=$$^{12}$C, $^{27}$Al, $^{40}$Ca, $^{93}$Nb, $^{208}$Pb, and $^{238}$U) collisions calculated
for incident proton kinetic energy of $T_0=14$ GeV on the basis of equations (6) and (7) for different
values of the genuine charmonium--nucleon absorption cross section $\sigma_{J/\Psi N}$, as indicated in
the inset, and for no $J/\Psi$ mass shift. While the calculations with only $l_{J/\Psi}=0$ in equation (7)
are given in the figure for $\sigma_{J/\Psi N}=7$ mb and $\sigma_{J/\Psi N}=14$ mb, for the case of
$\sigma_{J/\Psi N}=3.5$ mb the ones are presented here already for two options, namely: i) $l_{J/\Psi}=0$
and ii) $l_{J/\Psi}=0.8$ fm. A choice of  $\sigma_{J/\Psi N}=3.5$ mb has been particularly motivated by the
results from the $J/\Psi$ photoproduction experiment at SLAC [35, 36], while the value of
$\sigma_{J/\Psi N}=7$ mb
\footnote{$^)$It should be mentioned that the $J/\Psi N$ inelastic cross section of the order of
6--8 mb is reported in recent calculations [37] adopting effective Lagrangians.}$^)$
was dictated by the analyses of the observed suppression of $J/\Psi$ in nuclear collisions within the
various models [1, 5, 6] based on $J/\Psi$ absorption by hadrons. Finally, an option of
$\sigma_{J/\Psi N}=14$ mb was motivated by the findings of [11], indicating that the charmonium--nucleon
absorption cross section can be larger than $\approx$ 10 mb at FAIR energies
\footnote{$^)$It is interesting to note in this connection that the cross section of the $J/\Psi$
--nucleon elastic scattering at the threshold was found in [27] to be most likely larger than 17 mb.}$^)$.
\begin{figure}[htb]
\begin{center}
\includegraphics[width=12.0cm]{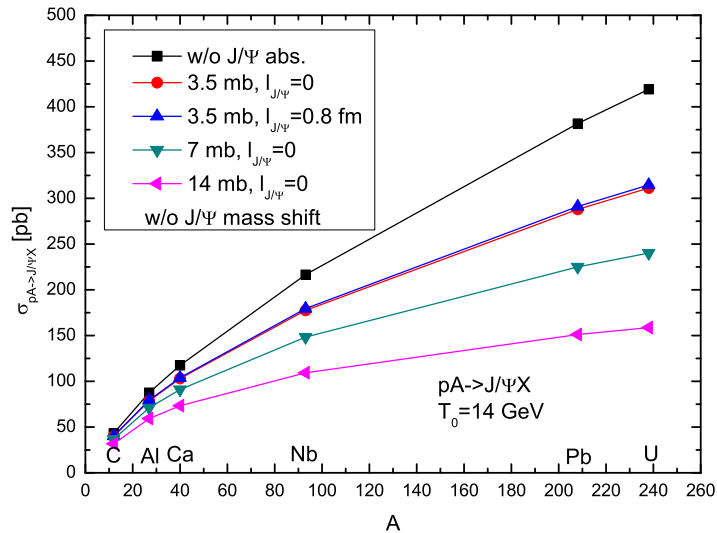}
\vspace*{-2mm} \caption{A--dependence of the total cross section of $J/\Psi$ production
by 14 GeV protons from primary $pN \to pN{J/\Psi}$ channels in the full phase space in the scenario without $J/\Psi$
mass shift for different values of the $J/\Psi N$ absorption cross section and $J/\Psi$ formation length indicated
in the inset.
The lines are included to guide the eyes.}
\label{void}
\end{center}
\end{figure}
One can see that the results are practically insensitive to the $J/\Psi$ formation time effects.
On the other hand, they depend strongly, especially for heavy target nuclei, on the charmonium--nucleon
absorption cross section. Looking at this figure, we see also that for the incoming proton energy of 14 GeV
of our interest the value of the absolute $J/\Psi$ meson yield is of the order of 100--400 pb for targets
heavier than the Al target in employed four scenarios for the cross section $\sigma_{J/\Psi N}$. This value
is very small, but one might expect to measure it in the future FAIR experiments. Therefore, we can conclude
that the observation of the A dependence, like that just considered, can serve as an important tool to determine
the genuine $J/\Psi N$ absorption cross section.

In figure 2 we show our predictions for the transparency ratio $T_A$, defined by equation (15) and calculated
using the results presented in the above figure, as a function of the nuclear mass number A for initial proton
energy of 14 GeV. It can be seen that, contrary to the preceding case, there are no differences between the results
obtained by adopting different $J/\Psi$ formation lengths under consideration. Whereas, we may observe in this
figure the experimentally separated differences ({$\sim$}20--30\%) between all calculations corresponding to different
options for the $J/\Psi N$ absorption cross section only for targets heavier than the Ca target, where they are
less than {$\sim$}10\%, which means that this observable is well suited to determine the cross section
$\sigma_{J/\Psi N}$ and
the future data from the CBM experiment for it performed using the heavy targets
(Nb, Pb) should help to distinguish between these options.
\begin{figure}[!h]
\begin{center}
\includegraphics[width=12.0cm]{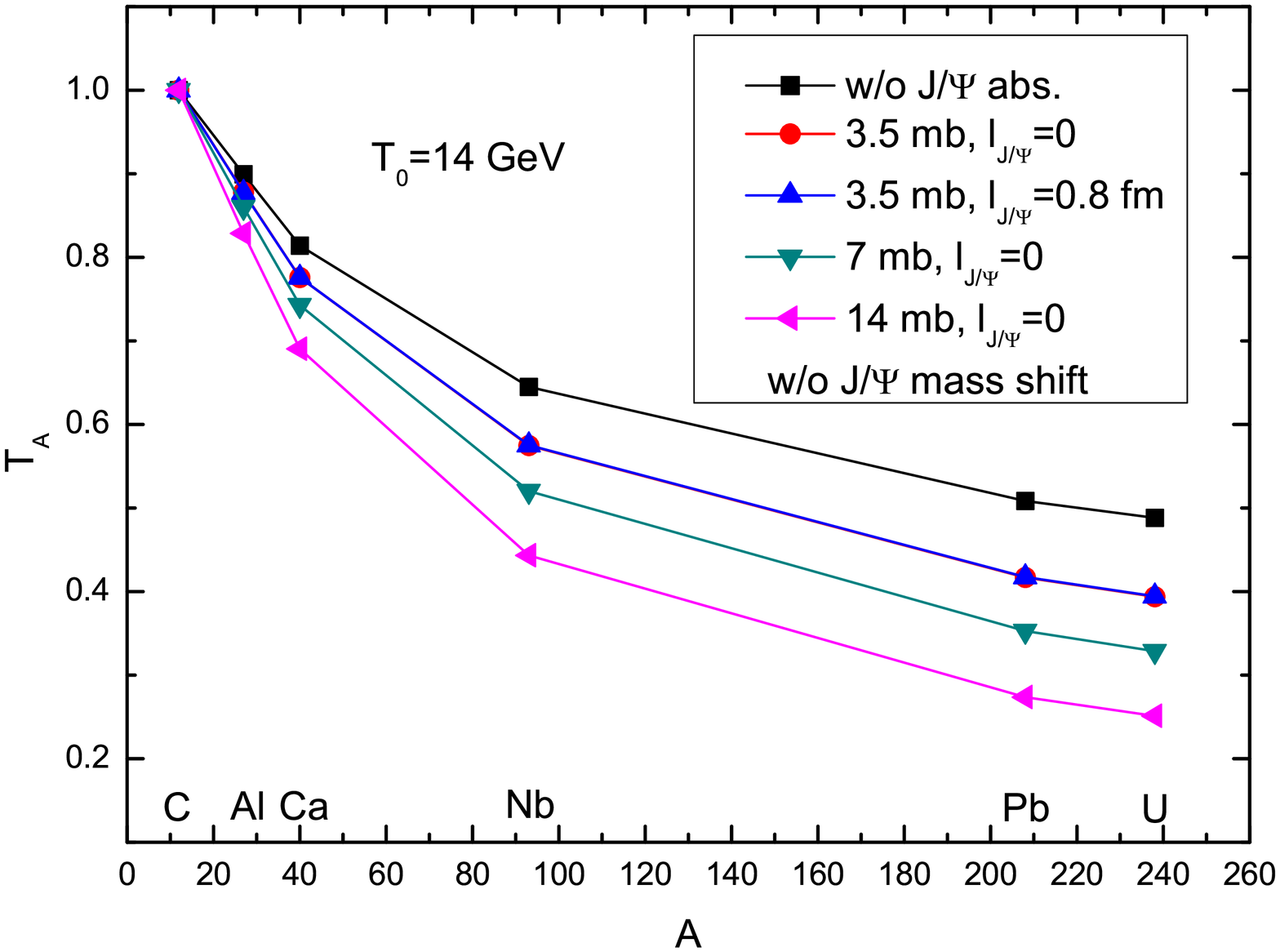}
\vspace*{-2mm} \caption{Transparency ratio $T_A$ for $J/\Psi$ mesons as a function of the
nuclear mass number $A$ in the scenario without their mass shift as well as for their different absorption
cross sections and formation lengths indicated in the inset. The lines are included to guide the eyes.}
\label{void}
\end{center}
\end{figure}
\begin{figure}[!h]
\begin{center}
\includegraphics[width=12.0cm]{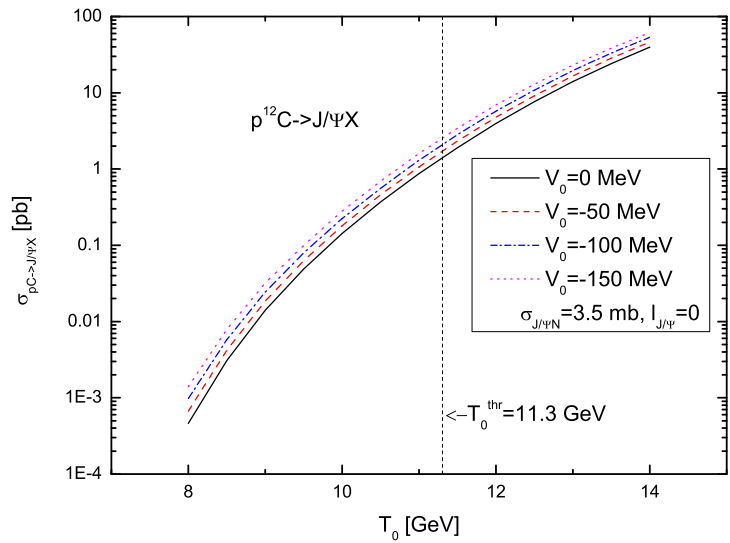}
\vspace*{-2mm} \caption{Excitation function for production of $J/\Psi$
mesons off $^{12}$C. The curves are calculations for $\sigma_{J/\Psi N}=3.5$ mb and $l_{J/\Psi}=0$ with
an in-medium $J/\Psi$ mass shift depicted in the inset. The vertical dashed line indicates the threshold energy
for $J/\Psi$ production on a free nucleon.}
\label{void}
\end{center}
\end{figure}
\begin{figure}[!h]
\begin{center}
\includegraphics[width=12.0cm]{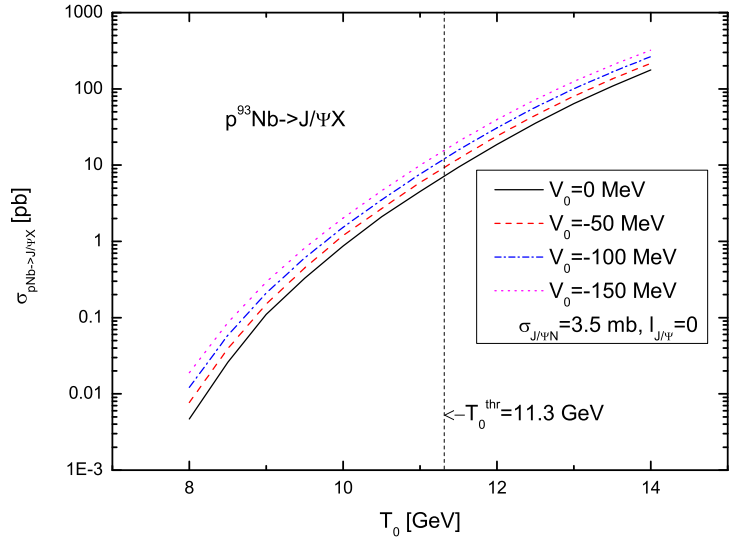}
\vspace*{-2mm} \caption{The same as in figure 3, but for the $^{93}$Nb target nucleus.}
\label{void}
\end{center}
\end{figure}

    Finally, we consider the excitation functions for production of $J/\Psi$ mesons off
$^{12}$C and $^{93}$Nb target nuclei. They were calculated on the basis of equation (6) for
$\sigma_{J/\Psi N}=3.5$ mb and $l_{J/\Psi}=0$ as well as for four adopted
scenarios for the $J/\Psi$ in-medium mass shift, and are given in figures 3 and 4.
One can see that in the far subthreshold region ($T_0 \sim$8--10 GeV) there are well separated predictions for
these considered scenarios for the $J/\Psi$ in-medium mass shift. The values of the total charmonium production
cross sections in this region are very small (in the range of 0.01--1 pb), but one might expect to measure
their in the future FAIR experiments as well.
Therefore, these measurements might help to get definite
information about this shift. It should be noticed that an analogous possibility has been
discussed before for the photoproduced $\omega$ [38] and $\eta^\prime$ [31] mesons, and it was very recently
realized for the $\eta^\prime$ mesons in [39].

   Taking into account the above considerations, we come to the conclusion that such observables as
the absolute and relative (transparency ratio) $J/\Psi$ meson yields from $pA$ interactions as
well as its excitation function can be useful at proton beam energies close to the kinematic threshold
to help determine both the genuine $J/\Psi N$ absorption cross section and
a possible charmonium mass shift in cold nuclear matter.

\section*{4. Conclusions}

\hspace{1.5cm} In this paper we have calculated the A dependence of the absolute and relative cross sections for $J/\Psi$ production from $pA$ collisions at 14 GeV beam energy by considering incoherent primary proton--nucleon
charmonium production processes in the framework of a nuclear spectral function approach, which accounts for
the struck target nucleon momentum and removal energy distribution, elementary cross section for proton--nucleon
reaction channel close to threshold as well as different scenarios for the genuine $J/\Psi N$ absorption cross section
and its formation length. Also we have calculated the excitation function for $J/\Psi$ production off $^{12}$C
and $^{93}$Nb target nuclei in the near-threshold energy regime.
It was found that the A dependence of the absolute and relative $J/\Psi$ yields at incident energy of interest,
on the one hand, is practically not influenced by formation length effects and, on the other hand, it is
appreciably sensitive to the charmonium--nucleon absorption cross section. This gives a nice opportunity to
determine it experimentally. It was also shown that the excitation function for
$J/\Psi$ production off nuclei is well sensitive to the possible $J/\Psi$ in-medium mass shift
at subthreshold beam energies, and  this offers the possibility to investigate the shift via $J/\Psi$
production on light and heavy target nuclei at these energies.
\\

\end{document}